# Mechanically Modulated Tunneling Resistance in Monolayer MoS$_2$

Deyi Fu,[1] Jian Zhou,[1] Sefaattin Tongay,[1] Kai Liu,[1,2] Wen Fan,[1] Tsu-Jae King Liu[3] and Junqiao Wu[1,2*]

[1]*Department of Materials Science and Engineering, University of California, Berkeley, CA 94720, USA*

[2]*Division of Materials Sciences, Lawrence Berkeley National Laboratory, Berkeley, CA 94720, USA*

3 *Department of Electrical Engineering and Computer Science, University of California, Berkeley, CA 94720, USA*

[*]E-mail: wuj@berkeley.edu

**Abstract**

We report on the modulation of tunneling resistance in MoS$_2$ monolayers by nano-indentation using an atomic force microscope (AFM). The resistance between the conductive AFM tip and the bottom electrode separated by a monolayer MoS$_2$ is reversibly reduced by up to 4 orders of magnitude, which is attributed to enhanced quantum tunneling when the monolayer is compressed by the tip force. Under the WKB approximation, the experimental data is quantitatively explained by using the metal-insulator-metal tunneling diode model. As an ideal tunneling medium, the defect-free, nanometer-thick MoS$_2$ monolayer can serve as the active layer for non-impacting nano-electro-mechanical switches.

With continuous down scaling of transistor size driven by the Moore's law, fundamental limits of the field effect transistor have resulted in a power density crisis for integrated circuit chips. This is because the transistor operating voltage has not been proportionally reduced in recent technology generations, due to the non-scalability of the threshold voltage. The minimum threshold voltage for a given off-state leakage current specification is set by the sub-threshold swing, which is constrained by Boltzmann statistics to be no less than 60 mV/decade at room temperature, and which ultimately limits the energy efficiency of CMOS technology.[1] To circumvent this limit in order to continue to advance information technology, nano-electro-mechanical (NEM) switches have been proposed as an alternative technology for future ultralow-power digital integrated circuits on the basis of two apparent advantages: zero OFF-state current and abrupt ON/OFF switching, which give rise to zero standby power consumption and extremely steep switching (< 0.1 mV/decade), respectively.[2] However, conventional NEM switch designs operate by making and breaking physical contact between the conducting electrodes to turn the device ON and OFF, and therefore are inherently susceptible to reliability issues such as contact material degradation, surface adhesion, etc.[3,4] In order to lessen or eliminate these issues, a novel NEM switch without impacting parts, based on piezoelectric transduction, recently has been proposed by IBM researchers.[5] The channel of the switch is composed of a piezoresistive material which undergoes pressure-induced metal-insulator transition (MIT) upon compression by an expanding piezoelectric in response to an applied gate voltage. It is noted that piezoresistive behavior exists in different materials and structures, and can be triggered not only by stress-induced MIT, but also by other mechanisms such as mechanically controlled quantum tunneling, as in a tunneling NEM (tNEM) switch.[6] For tNEM switch applications, the piezoresistive material needs to be sufficiently thin (on the order of a few nanometers or less) in order for significant tunneling to occur, and be free of point defects, dislocations or grain boundaries so as to minimize the OFF-state leakage current. These requirements impose serious constrains on the choice of tunneling materials. For example, $SiO_2$ can be deposited with ultra-thin thickness, but then current leakage becomes significant.[7]

Recently, the transition-metal dichalcogenide semiconductor $MoS_2$ has attracted great interest because of its unusual electronic properties.[8] A bulk crystal of $MoS_2$ comprises S-Mo-S layers held together by van der Waals forces; each S-Mo-S layer (referred to as a monolayer) consists of two planes of S atoms and an intermediate plane of Mo atoms covalently bonded to each other (Fig. 1(a)). Because of the relatively weak inter-layer interaction and the strong intra-layer bonding, ultrathin, defect-free crystals of $MoS_2$ down to a single monolayer (~ 0.65 nm) can be formed by mechanical exfoliation.[9] The fact that thusly-obtained $MoS_2$ monolayers are semiconducting, ultra-thin, and nearly defect-free makes them an ideal material for implementing tNEM switches. In this work, we demonstrate mechanical modulation of tunneling resistance through monolayer $MoS_2$, and discuss the possibility of using it as the active tunneling material in non-impacting tNEM switches.

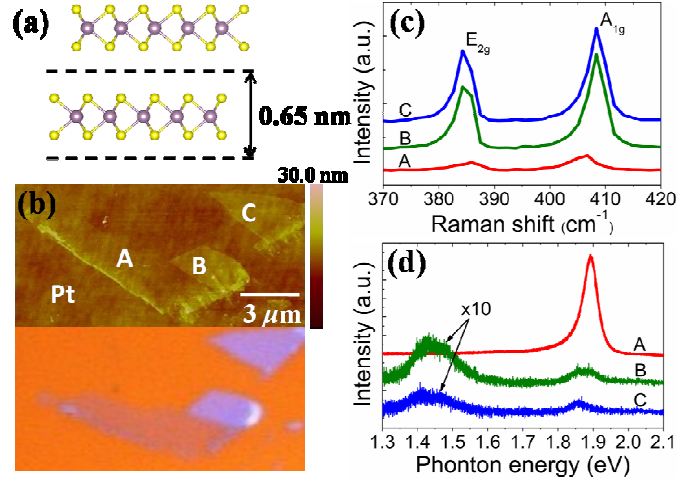

FIG. 1. (a) Covalently bonded S-Mo-S atoms in MoS$_2$ monolayers separated by van der Waals gap. (b) Upper panel: AFM image of MoS$_2$ flakes mechanically exfoliated on a Pt surface containing monolayer (A) and multilayer (B and C) regions. Lower panel: optical image of the flakes. (c) Raman spectra and (d) PL spectra of MoS$_2$ measured at different regions as shown in (b). The PL intensity of region B and C is magnified by 10 times and the curves in (c) and (d) are vertically offset for clarity.

Monolayer and multilayer MoS$_2$ flakes were mechanically exfoliated from bulk MoS$_2$ crystals onto Pt-coated 90 nm SiO$_2$/ Si substrates. Prior to the deposition of the Pt layer, a thin Cr layer was deposited by electron beam evaporation as an adhesive layer. The Pt/Cr bi-layer serves as bottom electrode in the following electrical measurements (Fig. 2(a)). The metal layers were thin (1 nm Cr and 2 nm Pt), so that the optical contrast between MoS$_2$ layers is preserved, and monolayer MoS$_2$ can still be optically located (lower panel of Fig. 1(b)). The upper panel of Fig. 1(b) shows an atomic force microscope (AFM) image of exfoliated MoS$_2$ flakes with both monolayer ("A") and multilayer ("B" and "C") regions. The multilayer flakes show characteristic A$_{1g}$ (out-of-plane) and E$_{2g}$ (in-plane) Raman modes located at around 408.4cm$^{-1}$ and 384.3cm$^{-1}$ respectively (Fig. 1(c)). For monolayer flakes, the A$_{1g}$ mode softens to 406.7 cm$^{-1}$ while the E$_{2g}$ mode stiffens to 385.9 cm$^{-1}$, which is consistent with previously reported observation.[10] It has been reported that MoS$_2$ undergoes a progressive crossover from indirect bandgap of ~ 1.29 eV in bulk crystal to direct bandgap of ~ 1.88 eV in a monolayer.[8] As shown in Fig. 1(d), the monolayer $MoS_2$ indeed shows strong band-edge photoluminescence (PL) at 1.88 eV, while for multilayer $MoS_2$, the direct-bandgap PL is drastically suppressed and a new PL peak emerges around 1.43 eV. The number of layers in the multilayer region is estimated to be 3 ~ 4.

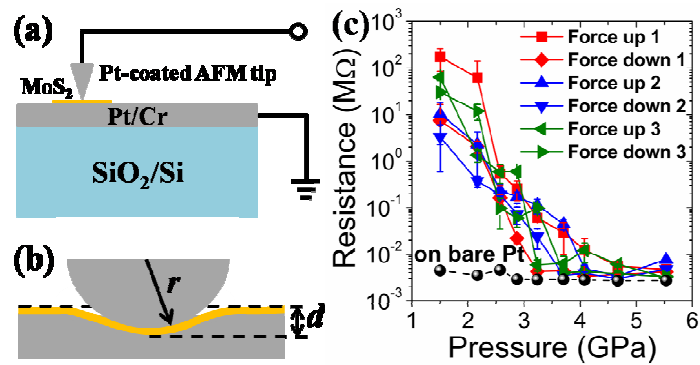

FIG. 2. (a) Schematic of the conductive-AFM measurement set up. (b) Sketch of Hertzian contact model showing an elastic sphere of radius $r$ indenting an elastic half-space to depth $d$. (c) Measured resistance vs. applied pressure. The contact resistance on the bare Pt surface is also shown for comparison.

Next we measure the vertical tunneling resistance of the monolayer $MoS_2$ using conductive-AFM as shown schematically in Fig. 2(a). A Pt-coated conductive AFM probe is used as the top electrode and indents the $MoS_2$ flake in the contact mode. This configuration allows us to apply compressive force directly onto the $MoS_2$ flake while simultaneously recording the current-voltage (I-V) relationship. By calibrating the spring constant of the AFM probe, we can accurately control the applied force. At each force point, the probe is stabilized for 1 min, and then an I-V curve is swept three times to check stability. The maximum bias voltage we applied is small (~ 50 mV) that the I-V curve is in the linear regime, therefore the resistance can be defined from the linear slope of the I-V curve near the origin. In order to quantify the pressure exerted onto the MoS$_2$ flakes, we employed the Hertzian contact model as depicted in Fig. 2(b). The model shows that an elastic sphere of radius $r$ indents an elastic half-space to depth $d$, and creates a contact area of $A=2\pi rd$. Here $d$ is related to the applied force by $F = \frac{4}{3}E^* r^{1/2} d^{/2}$, where $E^*$ is the reduced Young's modulus defined as $\frac{1}{E^*} = \frac{1-v_1^2}{E_1} + \frac{1-v_2^2}{E_2}$, $E_1$, $E_2$ are the Young's moduli, and $v_1$, $v_2$ the Poisson's ratios associated with each contact material (Pt and MoS$_2$). In Fig. 2(c), we show the measured resistance of the monolayer MoS$_2$ flake as a function of the calculated pressure. In order to confirm reproducibility of the measurements, we ramped the force up and down for multiple cycles. The resistance changes by ~ 4 orders of magnitude, and is highly reversible within the experimental sensitivity limited by instability factors such as thermal drift of the stage. It should be pointed out that this large ON/OFF ratio cannot be explained solely by contact resistance change between the probe and MoS$_2$ flake, because the estimated contact area varies only by a factor of 10 in the applied pressure range. We also note that the stress-induced resistance change is exponential, which is indicative of quantum tunneling process as the thickness of the monolayer MoS$_2$ (~ 0.65 nm) is modulated by the tip force. The typical definition of the atomic-scale thickness of monolayer MoS$_2$ actually takes into account the three atomic planes (S-Mo-S) plus the van-der-Waals gap (Fig. 1(a)), therefore its effective thickness could be modulated under pressure by shrinking the van der Waals gap as well as deforming the S-Mo-S bonding angles. Initially the monolayer MoS$_2$ deforms conformally with the bottom Pt substrate when the AFM probe indents it. Afterwards, the pressure exerted on the MoS$_2$ increases with the applied tip force, resulting in a decrease of the effective thickness of the monolayer MoS$_2$ separating the AFM tip and the bottom electrode. The resultant enhancement in quantum tunneling lowers the system's resistance to a point that is limited by the intrinsic contact resistance in the system. When the applied force is gradually removed, the system relaxes back and the resistance increases due to reduced tunneling.

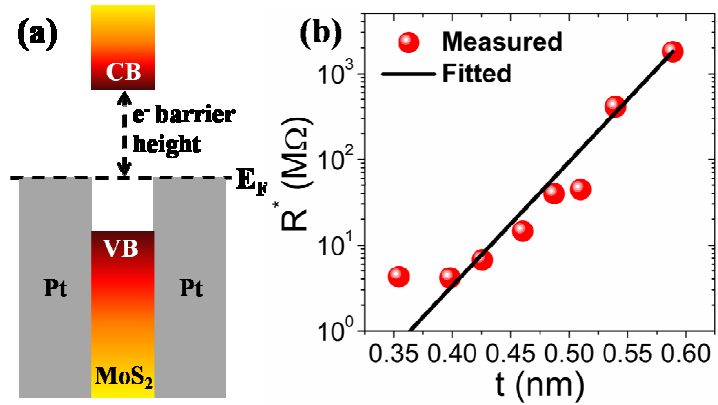

FIG. 3. (a) Band diagram of monolayer MoS$_2$ sandwiched between two Pt surfaces under equilibrium condition. (b) Effective resistance vs. barrier thickness as converted from Fig. 2(c).

When MoS$_2$ is in contact with metal, charge transfer occurs, which depletes the MoS$_2$ of mobile charge carriers. The band diagram is illustrated in Fig. 3(a), where the equilibrium Fermi level falls inside the bandgap of MoS$_2$, forming a typical symmetric metal-insulator-metal (MIM) tunneling diode structure. Depending on the metal quality and contact condition, the work function of Pt varies between 5.1 eV and 5.9 eV.[11] Assuming that the monolayer MoS$_2$ has the same electron affinity as that of bulk MoS$_2$ (~ 4 eV),[12] the electrons on the two sides experience a tunneling barrier of 1.1 ~ 1.9 eV. Under the WKB approximation,[13] the tunneling probability of charge carriers through a potential barrier height of $qV_b$ is: $T = e^{-2t\sqrt{2m^*V_b/\hbar^2}}$, where $t$ is the barrier thickness and $m^*$ is the carrier effective mass. When applied bias voltage is smaller than the barrier height, i.e. $V < V_b$, the tunneling current density of the MIM structure can be derived as[13]

$$J = J_0 \left[ (V_b - V/2) \exp\left(-C\sqrt{V_b - V/2}\right) - (V_b + V/2) \exp\left(-C\sqrt{V_b + V/2}\right) \right], \quad (1)$$

where $J_0 \propto t^{-2}$ and $C = 2t\sqrt{2m^*q/\hbar^2}$. The first term describes the tunneling current from cathode to anode while the second term describes the tunneling current in the reverse direction. In our case, $V \ll V_b$, Eq. (1) can be further approximated as

$$J = -J_0 V e^{-C\sqrt{V_b}}, \quad (2)$$

which shows a linear relationship between current and voltage. For a given contact area of $A$, the tunneling resistance can then be defined as

$$R = V/JA = e^{C\sqrt{V_b}}/J_0 A. \quad (3)$$

Since both contact area $A$ and barrier thickness $t$ change with applied force under the Hertzian contact model, it is more convenient to define an effective tunneling resistance as $R^* = R \cdot A/t^2 = R_0 e^{\alpha t}$, where $R_0$ is a constant regardless of $A$ and $t$, and the coefficient $\alpha$ is defined as $\alpha = 2\sqrt{2m^*qV_b/\hbar^2}$. For free electrons on the two sides of the barrier (monolayer MoS$_2$), they have the effective mass of $m_0$, therefore one can calculate the coefficient $\alpha$ to be in the range of 10 to 14 nm$^{-1}$. To our best knowledge, there is no report on the out-of-plane Young's modulus of monolayer MoS$_2$. If we follow the trend of graphene and assume the out-of-plane Young's modulus $E$ of monolayer MoS$_2$ to be 3.5% of the in-plane $E$,[14,15] Fig. 2(c) can be re-plotted in Fig. 3(b), where the data points represent the average of the values shown in Fig. 2(c). Fitting the equation $R^* = R_0 e^{\alpha t}$ to the linear part of the experimental data yields $\alpha$ = 33 nm$^{-1}$, which is larger than the estimated value, probably caused by uncertainties in the Young's modulus, tip radius, and electron affinity of monolayers.[16]

To summarize, we have measured the stress-induced resistance change of monolayer MoS$_2$ under the conductive-AFM configuration. The sharp AFM probe allows us to apply large stress of a few GPa, under which the resistance of monolayer MoS$_2$ is reversibly and exponentially modulated up to 4 orders of magnitude. The high ON/OFF ratio is attributed to quantum tunneling when the thickness of the monolayer MoS$_2$ is modulated by the force exerted by the probe tip. Under the WKB approximation, the experimental data is explained in the framework of MIM tunneling diode model. Since mechanically exfoliated monolayer MoS$_2$ is semiconducting, defect-free and nanometer thick, it can serve as a natural tunneling medium in a non-impacting NEM switch design. Bi-layer and few-layer MoS$_2$ also showed stress-modulated resistance change of 2 ~ 3 orders (not shown here), but with higher ON-state resistance, therefore are less attractive than monolayer MoS$_2$ for this application. It should be noted that besides MoS$_2$, any high-quality 2D semiconductors (such as monolayer WS$_2$, MoSe$_2$, Mg(OH)$_2$, BN, etc.) can be used as the tunneling material in a tNEM switch, where the ON/OFF ratio and turn-on voltage are determined by their electrical conductivity and mechanical compressibility, respectively.

This work was supported by the NSF Center for Energy Efficient Electronics Science (NSF Award No. ECCS-0939514). We are grateful to Prof. Eli Yablonovitch for helpful discussions.